\documentclass[manuscript]{acmart}
\usepackage{makecell}

\usepackage{graphicx}
\graphicspath{ {./images/} }
\AtBeginDocument{%
  \providecommand\BibTeX{{%
    \normalfont B\kern-0.5em{\scshape i\kern-0.25em b}\kern-0.8em\TeX}}}

\setcopyright{none}
\copyrightyear{2022}
\acmYear{2022}
\acmDOI{...}

\acmPrice{15.00}
\acmISBN{978-1-4503-XXXX-X/18/06}

\begin{document}

% NutFit: The effects of Nudging on Tracking Fitness
\title{The Nudging Effect on Tracking Activity }
% Impact on the Consistency of Wearing Activity Trackers

\author{Ruochun Wang}
\email{rwan8874@uni.sydney.edu.au}
\orcid{ 0000-0002-1097-1906}
\affiliation{%
  \institution{The University of Sydney}
  \city{Sydney}
  \state{NSW}
  \country{Australia}
  \postcode{2000}
}

\author{Amani Abusafia}
\email{amani.abusafia@sydney.edu.au}
\orcid{0000-0001-9159-6214}
\affiliation{%
  \institution{The University of Sydney}
  \city{Sydney}
  \state{NSW}
  \country{Australia}
  \postcode{2000}
}

\author{Abdallah Lakhdari}
\email{abdallah.lakhdari@sydney.edu.au}
\orcid{0000-0001-8005-1534}
\affiliation{%
  \institution{The University of Sydney}
  \city{Sydney}
  \state{NSW}
  \country{Australia}
  \postcode{2000}
}

\author{Athman Bouguettaya}
\email{athman.bouguettaya@sydney.edu.au}
\orcid{0000-0003-1254-8092}
\affiliation{%
  \institution{The University of Sydney}
  \city{Sydney}
  \state{NSW}
  \country{Australia}
  \postcode{2000}
}

\renewcommand{\shortauthors}{Ruochun and Amani, et al.}

\begin{abstract}
Wearables activity trackers are becoming widely adopted to understand individual behavior. Understanding behavior may help in self-regulation such as self-monitoring, goal-setting, self-corrective, etc.; Nevertheless,  challenges exist in attaining consistent use and adoption of wearables, which hinders behavior understanding. Research has suggested that nudging strategies may change and sustain human engagement.   However, it is still unknown how nudging may affect human wearing behavior on an individual level.   We conducted a six-month study in which we tested several nudging techniques on the same participants. The preliminary results of our research show that participants perform better when a nudging strategy is applied. In addition, participants responded differently to different nudging techniques. Future research can focus on developing an individual-based nudging mechanism to encourage users to wear their devices consistently.\looseness=-1
\end{abstract}





\maketitle

%https://dl.acm.org/doi/pdf/10.1145/2971648.2971656
%https://www.researchgate.net/profile/Patrick-Shih-2/publication/268746784_Use_and_adoption_challenges_of_wearable_activity_trackers/links/550f39ac0cf2ac2905ae01a8/Use-and-adoption-challenges-of-wearable-activity-trackers.pdf

\section{Introduction}

Wearable activity trackers (i.e., Fitbit) have been used to report health-related activities \cite{gulotta2016fostering, shih2015use,nuss2021motivation}. The wearables' regular use will lead to consistent reporting, which may help understand the users' behavior. Understanding the behavior may help in several aspects, including activity recommendations, self-regulation, and health intervention \cite{epstein2016reconsidering}. However,  attaining consistent use and adoption of wearables is a challenge \cite{heatherton1994losing}. Indeed, people often tend to forget or stop wearing their devices over time \cite{epstein2016reconsidering}.   The inconsistency in using the wearables is a common challenge that hinders the full understanding of users' behavior \cite{shih2015use}.  A crucial question is how to change behaviors to increase and sustain regular activity tracking? 

Nudging is a key mechanism to increase and motivate regular human engagement \cite{caraban201923}. In this paper, we aim to study the impact of nudging strategies on individuals' adherence to wearing activity trackers. Even though several studies have explored the use of nudging theory to change human behavior \cite{caraban201923}. Most of these studies focused on changing instant behavior in the short term. In addition, the activity of wearing a tracker requires daily motivation and remembrance. Hence, it may require different nudging strategies. Additionally, most of these studies would try a different nudging strategy for each group. This suggests that people will respond to nudging strategies similarly \cite{caraban201923,tseng2018different}. However, in reality, individuals react differently based on their individual characteristics and preferences. Therefore, this paper aims to study the effect of several nudging strategies on each individual over time. For instance,  all participants were studied using the same nudging strategies. Furthermore, our study used different email reminders as a nudging tool reminding participants to wear and use their gadgets once every month. Then we assess the change in their device-wearing behavior in terms of the number of days they wore their devices.

%  What is the problem?
 
% The objective of this project is to conduct a long-term experiment to study the impact of different incentives on the consistency of the participants in wearing their Fitbit devices. Moreover, we aim to model the different motivation patterns of the participants. The model will reflect how the incentive methods could influence the participants' reporting habits as well as the overall crowdsourcing data collection. Our model will utilize the users' previous behaviour to detect the most effective incentives to use with them.
\vspace{-8pt}
\section{Study Method}
%or education, there are 10 Bachelor's degree, 14 High school (year 12), 3 Master's degree and 1 PhD. 
Overall, we have recruited 28 participants to wear the Fitbit smartwatch and report health data for around 6 months (gender:  female = 13, male = 15; age: M = 21.71, SD =3; BMI: M =  23.46, SD = 3.95; education: High diploma = 10, Bachelors = 11, Masters  = 3, PhD = 1). Our data was collected through the Fitbit smartwatch remote back-end database. Fitbit smartwatch tracks user health data such as average breathing rate, steps count, and consumption of calories to the remote server in real-time. At the beginning of this study, we provided all participants with a Fitbit smartwatch and requested them to wear it daily. We computed their active rate as the percentage of the days they wore the devices over a month. A device is considered worn if the number of steps is more than one. Based on their first-month reporting active rates (baseline period), 28 participants were evenly divided into the experimental and control groups. Starting from the second month, we sent short emails containing a nudging message to the experiment group at the end of each month. In total, there are five different nudging strategies that we have used in our emails: 1) Neutral - Which contains the basic monthly activity data without emotional bias; 2) Positive - Motivational message, which explains the positive impact of their consistency in wearing the devices on their health and our research ; 3) Negative - A message that informs the participants that third of them have not worn their devices the past 30 days; 4) Rank - A message which mentions each participant their rank among all participants in the previous month and how consistent where the top three participants; 5) Monetary - A message which reminds participants that a monetary award would be obtained at the end of the month for their participation.\looseness=-1

\vspace{-8pt}
\section{Results}
\begin{figure}[!t]
\centering
    \begin{minipage}{0.49\linewidth}
    \centering
    \setlength{\abovecaptionskip}{1pt}
    \setlength{\belowcaptionskip}{-15pt}
    \includegraphics[width=\linewidth]{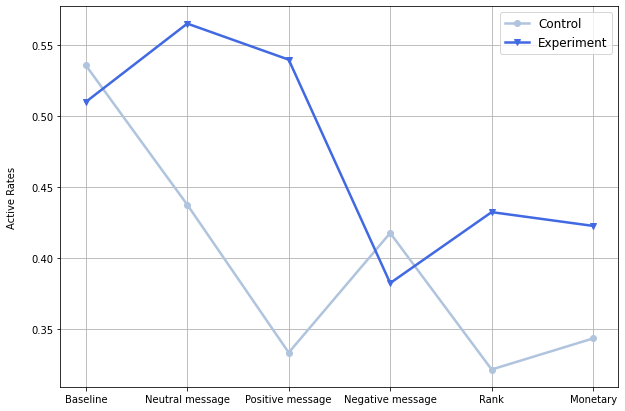}
    \caption{Active rates of both groups each month}
    \label{fig:active}
    \end{minipage}
    \hfill
    \begin{minipage}{.49\linewidth}
    \centering
    \setlength{\abovecaptionskip}{1pt}
    \setlength{\belowcaptionskip}{-15pt}
    \includegraphics[width=\linewidth]{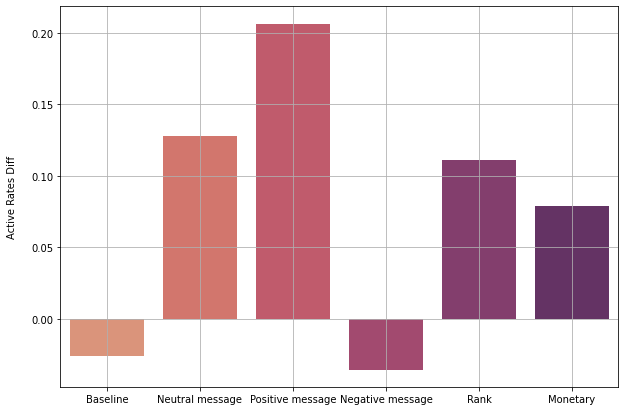}
    \caption{Average active rates difference between experiment and control groups each month }
    \label{fig:diff}
    \end{minipage}
\end{figure}

\begin{figure}[!t]
\centering
    \setlength{\abovecaptionskip}{1pt}
    \setlength{\belowcaptionskip}{-15pt}
    \includegraphics[width=\linewidth]{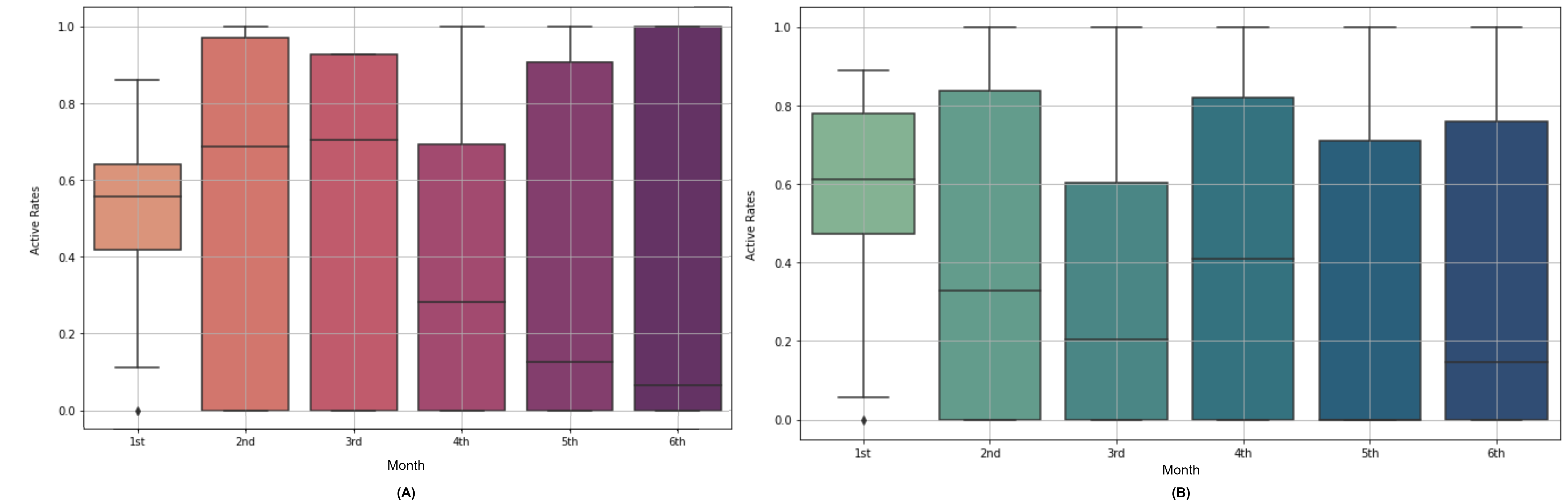}
    \caption{Active rates distribution in each month for (A) Experiment group (B) Control group}
    \label{fig:pplot}
\end{figure}

The first experiment studies the impact of each nudging strategy by comparing the average active rates between the experiment and the control groups in each month (See Fig.\ref{fig:active}). Note that the experiment group's participants who did not open their email in a month were dropped from that month. The preliminary results show that the experimental group had a higher active rate than the control group, except when the message was negative. According to the results, all nudging methods increased the active rates of participants by 5-20\% except for the negative message, which slightly reduced the participation by 3\% (See Fig.\ref{fig:diff}). Additionally, the results show that the positive message has the highest impact. Another observation is that, generally, the active rates of the participants of both groups have declined over time. This is due to the natural habit of losing motivation in humans as time passes \cite{epstein2016reconsidering}.

We analyze the participants' active rates distribution to ascertain the previous experiment. The previous experiment only considers the average active rate in each month. We visualize the distribution of both groups as shown in Fig.\ref{fig:pplot}. In general, both figures show a wide distribution of active rates across the months except for the first baseline month. This distribution indicates that  (1) Nudging techniques impact people differently, e.g., some will be more motivated by the positive messages while others will be motivated by ranking. This observation gives us an insight that individuals' behavior varies with nudging strategies. A customized nudging mechanism is needed to identify the right nudging for each individual based on their previous behavior. (2) people's behavior in consistently wearing their Fitbit is different; this is reflected in the active rates distribution of the control group (Fig.\ref{fig:pplot} (B)). Lastly, the distribution of the active rates in the first month is condensed. This may be justified by the excitement of participants at the start of the project \cite{epstein2016reconsidering}.

\vspace{-8pt}
\section{Conclusion}
We have shown that different nudging strategies impact people's consistency in wearing their activity trackers. Our future work will be to build a nudging mechanism that considers the individual behavior of participants. We will also recruit more participants to have an extensive study.

\vspace{-8pt}
\section*{Acknowledgment}
This research was partly made possible by  LE220100078 and LE180100158 grants from the Australian Research Council. The statements made herein are solely the responsibility of the authors.
\vspace{-8pt}

\bibliographystyle{ACM-Reference-Format}
\bibliography{main}

\appendix









\end{document}